\documentclass[onecolumn,amsfonts,showpacs,superscriptaddress]{revtex4-1}
\usepackage[dvipdfmx]{graphicx}
\usepackage{amssymb,amsmath,amsthm,booktabs,mathtools}
\usepackage{bbm}
\usepackage{bm}
\usepackage{color}
\usepackage[colorlinks,bookmarks=false,citecolor=blue,linkcolor=red,urlcolor=blue]{hyperref}
\usepackage{tikz}
\usepackage{pgfplots}
\usepackage{enumitem}

\def\tit#1{}

\usepackage{color}

\usepackage{epsfig}
\usepackage{amscd}
\usepackage{graphicx,xspace}
\input{epsf}

\include{latexcommands}

\def\be{\begin{equation}}
\def\ee{\end{equation}}

\def\bea{\begin{eqnarray}}
\def\eea{\end{eqnarray}}

\usepackage{amsmath}

\usepackage{amsmath}	
\begin{document}

\title{Quantum Generalized Hydrodynamics of the Tonks-Girardeau gas:\\ density fluctuations and entanglement entropy}

\author{Paola Ruggiero}
\affiliation
{Department of Quantum Matter Physics, University of Geneva, 24 Quai Ernest-Ansermet, CH-1211 Geneva, Switzerland}
\author{Pasquale Calabrese}
\affiliation
{SISSA and INFN, Via Bonomea 265, 34136 Trieste, Italy}
\affiliation
{International Centre for Theoretical Physics (ICTP), Strada Costiera 11, 34151 Trieste, Italy}
\author{Benjamin Doyon}
\affiliation
{Department of Mathematics, King’s College London, Strand WC2R 2LS, UK}
\author{J\'er\^ome Dubail}
\affiliation
{Laboratoire de Physique et Chimie Th\'eoriques, CNRS, UMR 7019, Universit\'e de Lorraine, 54506 Vandoeuvre-les-Nancy, France
}

\begin{abstract}
We apply the theory of Quantum Generalized Hydrodynamics (QGHD) introduced in [Phys. Rev. Lett. 124, 140603 (2020)] to derive asymptotically exact results for the density fluctuations and the entanglement entropy of a one-dimensional trapped Bose gas in the Tonks-Girardeau (TG) or hard-core limit, after a trap quench from a double well to a single well. 
On the analytical side, the quadratic nature of the theory of QGHD is complemented with the emerging conformal invariance at the TG point to fix the universal part of those quantities. Moreover, the well-known mapping of hard-core bosons to free fermions, allows to use a generalized form of the Fisher-Hartwig conjecture to fix the non-trivial spacetime dependence of the ultraviolet cutoff in the entanglement entropy.
The free nature of the TG gas also allows for more accurate results on the numerical side, where a higher number of particles as compared to the interacting case can be simulated. The agreement between analytical and numerical predictions is extremely good. For the density fluctuations, however, one has to average out large Friedel oscillations present in the numerics to recover such agreement.
\end{abstract}

\maketitle

\tableofcontents

\section{Introduction}

\subsection{Hydrodynamics, quantum fluctuations, and equal-time correlations}

Hydrodynamics offers a powerful way of thinking about the dynamics of many-body systems on macroscopic scales~\cite{landau1987course,spohn2012large}. While made of microscopic constituents that typically undergo complex collective dynamics, fluids of many particles are viewed as continuous media on larger scales. Their evolution is encoded in the time- and spatial-dependence of a small number of local thermodynamic quantities, or hydrodynamic variables, such as particle density, energy density, or more generally densities of conserved charges in the system. In that effective description at the macroscopic scale, the microscopic degrees of freedom are not immediately visible, but instead they are reflected in long wavelength variations of the charge densities. In particular, around an equilibrium configuration, small variations of the charge densities propagate through the fluid as sound waves. These sound waves are the relevant collective degrees of freedom at low energy.

The basic ideas of hydrodynamics apply to fluids made of classical or quantum constituents alike. For instance, the Euler equations for a Galilean-invariant one-dimensional (1D) fluid with conserved particle number and momentum typically read, in conservative form,
\begin{equation}
    \label{eq:euler_zeroT}
    \left\{ \begin{array}{rcl}
        \partial_t \rho + \partial_x (\rho u) & = & 0 \\
        \partial_t (m \rho u) +   \partial_x ( m \rho u^2  +  \mathcal{P}(\rho) )    & = & -  \rho \partial_x V .
    \end{array}  \right.
\end{equation}
Here $\rho(x,t)$ is the particle density, $u(x,t)$ is the mean fluid velocity, $m$ is the particles mass, $V(x)$ is an external potential that acts as a source term for the momentum density $m \rho u$ according to Newton's second law, and $\mathcal{P}(\rho)$ is the equilibrium pressure computed in the underlying microscopic model. This microscopic model can be either classical or quantum: the Euler equations (\ref{eq:euler_zeroT}) are applicable to a classical gas at constant temperature, for instance in contact with a thermostat; they are also applicable to quantum gases at zero temperature~\cite{abanov2006hydrodynamics}, in particular to the one-dimensional Bose gas with contact repulsion~\cite{doyon2017large,ruggiero2019conformal}. Apart from the specific function $\mathcal{P}(\rho) $ entering the momentum current in (\ref{eq:euler_zeroT}), which differs from one microscopic system to another, the form of the equations (\ref{eq:euler_zeroT}) is the same.

Differences between quantum and classical fluids arise at the level of their fluctuations. In a classical fluid at finite temperature, one expects the low-energy collective modes to have thermal fluctuations, and it is the goal of `fluctuating hydrodynamics' approaches to capture these, see e.g.~\cite{landau1992hydrodynamic,swift1977hydrodynamic,forster2018hydrodynamic,spohn2014nonlinear}. In a quantum fluid at zero temperature, there are no thermal fluctuations, but the low-energy collective modes typically have quantum fluctuations. This is well illustrated starting from the Euler equations (\ref{eq:euler_zeroT}). For simplicity, consider the ground state of the spatially homogeneous system ($V(x)=0$), with $\rho (x,t) = \rho_0$ and $u(x,t) = 0$. Linearizing the system (\ref{eq:euler_zeroT}) for small deviations $(\delta \rho (x,t), \delta u(x,t))$, one finds $ \partial_t \delta \rho + \rho  \partial_x \delta u   =  0$ and $ \partial_t \delta u + \frac{1}{m \rho} \frac{\partial \mathcal{P}}{\partial n} \partial_x \delta \rho    =  0 $, 
or equivalently
\begin{eqnarray}
    \label{eq:sound_wave2}
   \left( \frac{\partial}{\partial t}  -  \left( \begin{array}{cc}
        + v & 0 \\  0 & - v
    \end{array} \right) \frac{\partial}{\partial x} \right)  \left( \begin{array}{c}
       \pi \,  \delta \rho + K  \frac{m}{ \hbar} \, \delta  u  \\
         \pi \,  \delta \rho - K  \frac{m}{ \hbar} \, \delta  u
    \end{array} \right) \, = \, 0, \qquad  \quad  {\rm with} \quad  v := \sqrt{\frac{1}{m}\frac{\partial \mathcal{P}}{\partial \rho}},  \qquad  K := \frac{\pi \hbar \rho}{m v} .
\end{eqnarray}
Here $v$ is the sound velocity in the fluid, and $K$ is a dimensionless parameter. [$K$ is called the Luttinger parameter, and it is normalized such that $K = 1$ in the Tonks-Girardeau (hard core) limit of the 1D Bose gas with contact repulsion]. We see from Eq.~(\ref{eq:sound_wave2}) that there are right- and left-moving sound waves, corresponding to specific linear combinations of $\delta \rho$ and $\delta u$ parameterized by $K$, traveling at velocity $\pm v$. The sound waves are then used as the basic ingredient in a quantized theory of the fluid described by the Euler equations~(\ref{eq:euler_zeroT}). The basic idea is to look at $\delta \rho(x)$ and $\delta u(x)$ as operators $\delta \hat{\rho}(x)$, $\delta \hat{u}(x)$ in a quantum theory, and to impose the canonical commutation relations~\citep{landau1941theory},
\begin{equation}
    \label{eq:landau_canonical}
    \left[ \delta \hat{u} (x) , \delta \hat{\rho} (y) \right] \, = \,  \frac{\hbar}{i m} \delta'(x-y) ,
\end{equation}
and $\left[ \delta \hat{\rho} (x) , \delta \hat{\rho} (y) \right] = \left[ \delta \hat{u} (x) , \delta \hat{u}(y) \right] = 0$. To construct a Hamiltonian  for these quantum fluctuations, one imposes that the Heisenberg equations $\partial_t \delta \hat{\rho} = \frac{i}{\hbar} [\hat{H}, \delta \hat{\rho}]$ and $\partial_t \delta \hat{u} = \frac{i}{\hbar} [\hat{H}, \delta \hat{u}]$ coincide with the equations of motion (\ref{eq:sound_wave2}). This leads to
\begin{equation}
    \label{eq:Luttinger_ham}
    \hat{H} \, = \, \frac{\hbar v}{2} \int \left[ \frac{K}{\pi} \left( \frac{m}{\hbar} \delta \hat{u}(x) \right)^2 + \frac{\pi}{K} (\delta \hat{\rho}(x))^2 \right] dx ,
\end{equation}
which is the Hamiltonian of a Luttinger liquid~\cite{giamarchi2003quantum,cazalilla2004bosonizing,tsvelik2007quantum}. In conclusion, when one looks at quantum fluctuations of the collective modes (sound waves) of a standard Euler fluid in 1D, one arrives at the Luttinger liquid, which is the universal theory of 1D quantum hydrodynamics.

Physically, one consequence of the quantum fluctuations of the collective modes is that they induce equal-time correlations at different points in the fluid. For instance, the connected part of the density-density correlation in the ground state of the Hamiltonian~(\ref{eq:Luttinger_ham}) is ~\citep{giamarchi2003quantum,cazalilla2004bosonizing,tsvelik2007quantum}:
\begin{equation}
    \label{eq:rhorho}
    \left< \delta \hat{\rho} (x_1) \delta \hat{\rho} (x_2) \right>_{\rm conn.} \, = \, \frac{-K}{2 \pi^2 (x_1-x_2)^2} .
\end{equation}
{This is the leading power-law decay for the zero-temperature correlation of microscopic density observables in the model whose hydrodynamic equations are \eqref{eq:euler_zeroT}.}
Similarly, correlations of many other observables can be obtained from simple calculations within Luttinger liquid theory. We stress that the presence of such {power-law} correlation functions between observables at different macroscopic positions in the fluid is really a quantum effect. They are not {accounted for in the classical fluid theory (\ref{eq:euler_zeroT}), which, by linear response, only predicts nonzero correlations in space-time at the velocities of sound mode propagation, under the Euler scaling $\lim_{\lambda\to\infty} \lambda \langle \delta\hat\rho(\lambda x_1,\lambda t_1) \delta\hat\rho(\lambda x_2,\lambda t_2)\rangle_{\rm conn.}$ \cite{denardisCorrelations21} in which \eqref{eq:rhorho} vanishes; and they are stronger than correlations that would occur at finite temperatures due to thermal fluctuations, which decay exponentially in space.} This is even more obvious when one considers the entanglement between different pieces of the fluid. Classically there can be no entanglement. Yet, after quantum fluctuations have been incorporated in the theory, it makes sense to look at the entanglement entropy in the ground state of the Hamiltonian (\ref{eq:Luttinger_ham}), leading to the well-known result for a subsystem of length $\ell$~\cite{cc-04,cc-09},
\begin{equation}
    \label{eq:SvN}
    S(\ell) \, = \, \frac{1}{3} \log (\ell / \epsilon ) ,
\end{equation}
for some cutoff $\epsilon$. This shows that, despite the simplicity of the approach, quantizing the collective modes of the hydrodynamic equations (\ref{eq:euler_zeroT}) has the potential to reveal many features of the fluid that are truly quantum.

\vspace{0.5cm}

The classical Euler equations (\ref{eq:euler_zeroT}) describe a fluid with conserved particle number and momentum. When there are more conserved quantities, these two equations are complemented with more conservation equations, one for each additional conservation law. Remarkably, even when there are infinitely many conserved quantities, it is still possible to derive an Euler-scale description. This is `Generalized Hydrodynamics' (GHD), the hydrodynamic theory of 1D integrable systems introduced in 2016 in two very influential papers by Castro-Alvaredo et al.~\cite{castro2016emergent} and by Bertini et al.~\cite{bertini2016transport}.

In a recent paper, we asked the following question~\cite{ruggiero2020quantum}: What is the theory of quantum fluctuations around GHD? In other words: What happens if one mimics the derivation of quantum hydrodynamics above, replacing the standard Euler equations (\ref{eq:euler_zeroT}) that lead to a standard Luttinger liquid (\ref{eq:Luttinger_ham}), by the GHD equations of Refs.~\cite{castro2016emergent,bertini2016transport}?

\vspace{0.5cm}

The answer given in Ref.~\cite{ruggiero2020quantum} is that the theory of quantum fluctuations around GHD is a multi-component, time-dependent and spatially inhomogeneous Luttinger liquid, where excitations propagate as the linear sound waves of the standard GHD theory. This setup was dubbed `quantum GHD' in Ref.~\cite{ruggiero2020quantum}. [We note that some authors have expressed their preference for other names for that setup, e.g. for `generalized quantum hydrodynamics'~\cite{alba2021reviewGHDinhomQuenches}. This terminology also makes sense, but to make the connection with our previous work~\cite{ruggiero2020quantum} perfectly clear, we keep the name `quantum GHD' in this paper.] Numerical comparisons of that theory with t-DMRG simulations for the 1D Bose gas at finite repulsion strength were presented in Ref.~\cite{ruggiero2020quantum}. In this follow-up paper, our goal is to investigate the special case of infinite repulsion (Tonks-Girardeau limit), which maps to non-interacting fermions~\cite{girardeau1960relationship}, so it allows to do more analytical calculations and to perform more stringent tests of the theory. We focus in particular on spatial- and time-dependence of density correlations, and on the evolution of the entanglement entropy in the system. We obtain a number of analytical results that are exact in the hydrodynamic limit, and for which we provide extensive numerical checks (Fig.~\ref{movie}). {We stress that as the Tonks-Girardeau limit is described by a free fermionic theory, whose hydrodynamic equations are linear, certain subtle correlation effects due to nonlinearities, discussed in \cite{ruggiero2020quantum}, are not present.} {Although our techniques are, to some extent, similar to semiclassics and to other recent related works~\cite{dean2018wigner,dean2019nonequilibrium,smith2020noninteracting,smith2021full}, to our knowledge there exists no alternative method to arrive at our results.}

\subsection{This paper: tests of `quantum GHD' in the Tonks-Girardeau gas}
\label{sec:model}

This paper is a follow-up of Ref.~\cite{ruggiero2020quantum}; we aim at clarifying some aspects of the results of~\cite{ruggiero2020quantum} by analyzing the simple case of the Tonks-Girardeau gas in more details. We start from the 1d Bose gas with delta repulsion in an external potential $V(x,t)$, with the Hamiltonian~\cite{lieb1963exact,berezin1964schrodinger,korepin1997quantum}
\begin{equation} \label{model}
	\hat{H}(t) \, = \, \int dx  \left(  \frac{\hbar^2}{2}  ( \partial_x \hat{\Psi}^\dagger ) (\partial_x \hat{\Psi}) + (V(x, t)-\mu) \hat{\Psi}^\dagger \hat{\Psi}  +   \frac{g}{2}  \hat{\Psi}^{\dagger 2} \hat{\Psi}^2 \right) ,
\end{equation}
where $\hat{\Psi}^\dagger(x) $, $\hat{\Psi}(x)$ are operators that create/annihilate a boson at position $x$, and satisfy the canonical commutation rule $[\hat{\Psi}(x), \hat{\Psi}^\dagger(x')]= \delta(x-x')$. $\mu$ is the chemical potential, and we set the mass of the bosons to $m=1$. The hard core (or Tonks-Girardeau~\cite{girardeau1960relationship}) limit is given by
\begin{equation} \label{eq:TG}
	g \, \rightarrow \, + \infty. 
\end{equation}
In that limit, the Hamiltonian \eqref{model} maps to the one of non-interacting fermions through the non-local (Jordan-Wigner) transformation,
\begin{equation}
	\label{eq:JW}
	\hat{\Psi}_{F}^\dagger (x) \, = \, e^{i \pi \int_{y<x}  \hat{\rho} (y) dy}\, \hat{\Psi}^\dagger(x),
\end{equation}
where $\hat{\rho}(y) = \hat{\Psi}^\dagger(y) \hat{\Psi}(y) = \hat{\Psi}_F^\dagger(y) \hat{\Psi}_F(y)$ is the particle density operator, such that the fermion creation/annihilation modes satisfy the canonical anti-commutation relations $\{ \hat{\Psi}_F(x) , \hat{\Psi}_F^\dagger(x') \} = \delta(x-x')$. In terms of the fermions, the Hamiltonian (\ref{model}) in the limit (\ref{eq:TG}) is quadratic,
\begin{equation} \label{Hf}
	\hat{H} (t)  = \int dx \left(  \frac{\hbar^2}{2} (\partial_x \hat{\Psi}_F^\dagger )  (\partial_x \hat{\Psi}_F)  +  (V(x,t) - \mu) \hat{\Psi}_F^\dagger \hat{\Psi}_F  \right),
\end{equation}
which allows to perform many analytical calculations that are impossible away from the hard core limit (\ref{eq:TG}).

Similarly to Ref.~\cite{ruggiero2020quantum}, we focus on the following protocol. The system is initially in the ground state $\left| \psi_0 \right>$ of the Hamiltonian \eqref{model} with $V(x, t=0) = a_4 x^4 - a_2 x^2$, representing a double-well trapping potential, and fixed chemical potential $\mu$.
Then, it is let evolve with the same Hamiltonian, but after a sudden change of the trap to $V(x, t>0) = \omega^2 x^2/2$. {This protocol can in principle be realized experimentally in ultracold gases: 1D gases near the Tonks-Girardeau limit have been realized e.g. in Refs.~\cite{kinoshita2004observation,kinoshita2006quantum,wilson2020observation}, and the temperature can be extremely low so that the gas is initially very close to its ground state. In particular, the validity of the zero-entropy GHD description~\cite{doyon2017large} of the gas reviewed in Section~\ref{sec:QGHD_TG} has been established experimentally in Ref.~\cite{malvania2020generalized}. Quenches from double-well to harmonic potentials can also be realized experimentally, see e.g. Ref.~\cite{schemmer2019generalized}.}

As in previous references~\cite{ruggiero2019conformal,ruggiero2020quantum}, a convenient way of taking the hydrodynamic limit is to fix the potential $V(x,t)$ and the chemical potential $\mu$, and then send $\hbar $ to zero.
Indeed, in the local density approximation (LDA), the number of particles in the system at $t=0$ can be estimated to be
\begin{equation}
	N   \underset{({\rm LDA})}{ \simeq }  \frac{1}{\hbar} \int \frac{dx}{\pi} \sqrt{2  (\mu - V(x,t=0))} ,
\end{equation}
where the integration domain is the interval where $\mu - V(x,t=0) > 0$ (we assume that it is a single interval). So we see that taking $\hbar \rightarrow 0$ is equivalent to taking the number of particles
\begin{equation}
	N \, \propto 1/\hbar  \rightarrow \, + \infty .
\end{equation}
All the results in this paper are obtained in that limit. 
Throughout the paper, $\left< \cdot \right>$ is the expectation value in the initial state $\left| \psi_0 \right>$.

\subsection{Organization of the paper}

In Section~\ref{sec:QGHD_TG}, we specialize the derivation of quantum fluctuations around GHD of Ref.~\cite{ruggiero2020quantum} to the Tonks-Girardeau gas. This allows us to introduce the main notations and concepts, in particular the concept of multiple (or split) Fermi seas, and of the Fermi contour in phase space. There, we also clarify the main difference between this work and previous works~\cite{ruggiero2019conformal,scopa2021exact}. In Section~\ref{sec:results_density} we present our results for the correlations of density fluctuations, and in Section~\ref{sec:results_entanglement} the results for the entanglement entropy. For the entanglement entropy, we need to compute a non-universal contribution of order $O(1)$ when $N \rightarrow \infty$ that is necessary to make quantitative comparisons with numerical simulations; this is done by extending results by Jin and Korepin~\cite{jk-04} and Keating and Mezzadri~\cite{km-05,km-long} in Subsection~\ref{subsec:fisherhartwig}. Our main results are summarized in Fig.~\ref{movie}. The details of the numerical simulation are explained in an Appendix.

\section{Generalized hydrodynamics and its quantum fluctuations: the non-interacting case}
\label{sec:QGHD_TG}
In this section we review the generalized hydrodynamics description of the Tonks-Girardeau gas and its semi-classical quantization. For a more general discussion also including the general interacting case, see, e.g., the recent review~\cite{alba2021reviewGHDinhomQuenches}.

\subsection{`Generalized Hydrodynamics' of non-interacting fermions: free evolution of the Wigner function}
%
%
For the Tonks-Girardeau gas, GHD is nothing but the evolution equation for the Wigner function of the underlying non-interacting fermions (\ref{eq:JW})~\cite{doyon2017large,ruggiero2019conformal} . The Wigner function is defined as~\cite{wigner,hillery1984distribution,cahill1969density}
\begin{equation}
n(x,p,t) =  \int dy \, e^{i \frac{p y}{\hbar}} \left<  \hat{\Psi}^\dagger_F(x+y/2,t) \hat{\Psi}_F(x-y/2,t) \right> 
\end{equation}
and it satisfies the evolution equation
\begin{equation}
	\label{eq:wigner}
	\partial_t n  +  p \partial_x n =  (\partial_x V(x,t))  \partial_p n .
\end{equation}
This is the classical Liouville equation, or equivalently the GHD equation for non-interacting particles. It has been used extensively in the study of out-of-equilibrium 1D systems long before the advent of GHD, see e.g.~\cite{bettelheim2006orthogonality,bettelheim2008quantum,bettelheim2011universal,bettelheim2012quantum,protopopov2013dynamics}. Strictly speaking, Eq.~(\ref{eq:wigner}) is exact only if the potential $V(x)$ is harmonic. If it is not harmonic, then Eq.~(\ref{eq:wigner}) is the zeroth order in an $\hbar$-expansion of the true evolution equation, known as the Moyal evolution equation~\cite{moyal1949stochastic}. Recently, such corrections have been addressed in Refs.~\cite{fagotti2017higher,dean2018wigner,fagotti2020locally}. We discuss briefly these corrections in our Conclusion. But for the quench protocol we are interested in, the potential $V(x, t>0)$ is harmonic so we need not worry about them in this paper. \vspace{0.3cm}

Next, while in the free case \eqref{eq:wigner} is valid at the microscopic level, here we want to interpret $n (x, p, t)$ as a coarse grained (or semiclassical) slowly varying distribution function in position and momentum. A special role in what follows is played by the \emph{zero-entropy states}~\cite{doyon2017large}, defined as follows,
\begin{equation}
	\label{eq:wigner_gamma}
	n(x,p,t) = \left\{ \begin{array}{rcl}
				1 & \quad & {\rm if} \; (x,p) \;{\rm is \; inside \; the \; contour} \; \Gamma_t \\
				0 & \quad & {\rm if } \; (x,p) \; {\rm \; is \; outside} \; \Gamma_t .
				\end{array} \right.
\end{equation}
The zero-entropy states are fully specified by the contour $\Gamma_t$, a set of points $(x_t, p_t)$ in phase-space known as \emph{Fermi contour} (see also Eq.~\eqref{eq:parameter_s} below).
Moreover, locally (around a given $x$), they take the form of \emph{split Fermi seas}~\cite{fokkema2014split,eliens2016general,vlijm2016correlations,eliens2017quantum},
\begin{equation} \label{splitFermiseas}
	 n(x, p, t) = \left\{ \begin{array}{rcl}
				1 & \quad & \; p \in [p_{1} (x,t) ,p_{2}(x,t)] \cup \dots \cup [p_{2Q-1} (x,t),p_{2Q}(x,t)] , \\
				0 & \quad & {\rm otherwise} .
				\end{array} \right.
\end{equation}
where $p_{a} (x,t), \, a \in \{1, \cdots , 2Q \}$, denote the Fermi points at position $x$ and time $t$.
Analogous split Fermi seas can be defined also in the interacting case~\cite{fokkema2014split,eliens2016general,vlijm2016correlations,eliens2017quantum}.
For such states, GHD, namely Eq.~\eqref{eq:wigner}, which can be seen an infinite number of equations (one for each value of the momentum $p$), reduces to a finite number of equations, as many as the number of these Fermi points. Such equations take the form of Burgers' equations~\cite{ruggiero2019conformal,bettelheim2006orthogonality,bettelheim2008quantum,bettelheim2011universal,bettelheim2012quantum,protopopov2013dynamics}
\begin{equation}
	\label{eq:kF_evol}
	 \partial_t p_{a}(x,t)  + { p_{a}(x,t)} \partial_x p_{a}(x,t) =  -  \partial_x V(x,t)  .
\end{equation}
In the special case when there are only two Fermi points ($Q=1$), the two resulting equations can be interpreted as those of conventional Euler hydrodynamics at zero temperature, Eqs.~(\ref{eq:euler_zeroT}). Indeed, they are just a change of variables with respect to the more common equations for density and hydrodynamic velocity, see e.g. the discussion in Ref.~\cite{ruggiero2019conformal}. Conversely, if more than two Fermi points appears, GHD cannot be reduced to conventional hydrodynamics anymore~\cite{doyon2017large}.

It is sometimes convenient to rewrite Eq. (\ref{eq:kF_evol}) in the following form. Parameterizing the contour $\Gamma_t$ clockwise by a parameter $s$ from $0$ to $2\pi$, i.e.
\begin{equation} \label{eq:parameter_s}
\Gamma_t = \left\{ (x_t(s), p_t(s)) ; \, s \in [0,2\pi) \right\},
\end{equation}
the points $(x_t(s) , p_t(s))$ of the contour simply move according to Newton's equation,
\begin{equation}	\label{eq:xt_pt}
	\frac{d}{dt} \left( \begin{array}{c}
		x_t(s) \\
		p_t (s)
	\end{array} \right) \, = \, \left( \begin{array}{c} 
		p_t(s)  \\
		- \partial_x V (x_t(s))
	\end{array} \right) .
\end{equation}

\vspace{0.5cm}

\begin{figure*}[t]
\centering
\includegraphics[width = 0.7\textwidth]{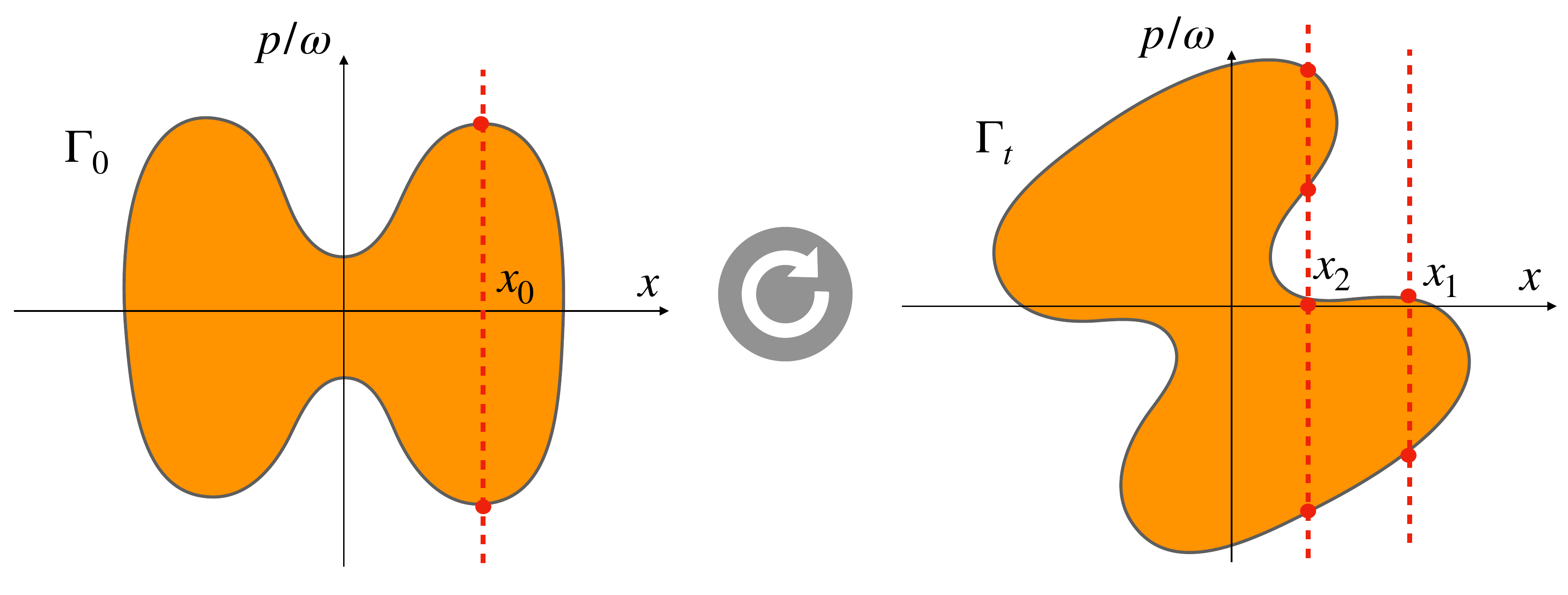}
\caption{Left: Wigner function at initial time $t=0$. For any choice of $x_0$ the maximum number of Fermi points (intersections between the vertical line $x=x_0$ and the contour $\Gamma_0$) is two. Right: Evolved Wigner function according to Eq.~\eqref{eq:wigner}: this corresponds to a simple rotation in the rescaled coordinates $(x, p/\omega)$ (cf. \eqref{eq:evo}). After time $t>0$ there are different regions, associated to a different number of Fermi points. For example, for a fluid cell around $x_1$ the number of Fermi points is still two. However, around $x_2$, there are four of them.}
\label{cartoon}
\end{figure*}

We now argue that the protocol introduced in Section~\ref{sec:model} can be described in terms of zero-entropy states, initially with two Fermi points ($Q=1$) and then at later times with four Fermi points ($Q=2$), see Fig.~\ref{cartoon}. Indeed, at initial time, the state of the system (in the large $N$ limit) is of the form~\eqref{eq:wigner_gamma}, and the Fermi contour $\Gamma_0$ has a ``butterfly-shape'' (Fig.~\ref{cartoon}, left). Specifically, it reads
\begin{equation}
n (x, p, 0) = \left\{ \begin{array}{rcl}
				1 & \quad  {\rm if} \; |p| \le \sqrt{2 (\mu - V(x, 0))} , \\
				0 & \quad  {\rm otherwise}  .
				\end{array} \right. 
\end{equation}
This means that locally it is parametrized  by a single pair of Fermi points ($Q=1$).
However, at time $t>0$ the evolution takes place within a harmonic trap $V(x,t>0)$. 
If $(x_0, p_0)$ is a given point of $\Gamma_0$, the corresponding point $(x_t, p_t)$ of the evolved contour $\Gamma_t$ at a given time $t$ is (cf. Eq.~\eqref{eq:xt_pt})
\begin{equation} \label{eq:evo}
\begin{pmatrix}
x_t \\
p_t/\omega
\end{pmatrix}
 =  
\begin{pmatrix}
\cos (\omega t) & \sin (\omega t) \\
-\sin ( \omega t) & \cos (\omega t)
\end{pmatrix}
\begin{pmatrix}
x_0 \\
p_0/\omega
\end{pmatrix},
\end{equation}
meaning that, with the rescaling $p_t \rightarrow p_t / \omega$, $\Gamma_t$ simply rotates in phase space at the trap frequency $\omega$. [In contrast with the interacting case discussed in Ref.~\cite{ruggiero2020quantum}, $\Gamma_t$ is not deformed under time evolution.]
Then, after some fraction of the period of the trap $\tau = 2\pi/\omega$, a region appears near the center with a split Fermi sea, $Q=2$ (Fig.~\ref{cartoon}, right). 

We stress that the appearance of multiple Fermi seas is a major difference with respect to the situations addressed in Refs.~\cite{ruggiero2019conformal,scopa2021exact}. While there the hydrodynamic problem is equivalent to a conventional form of hydrodynamics, the protocol considered here (as well as in Ref.~\cite{ruggiero2020quantum}) is the simplest generalization where GHD is really needed (see also Ref.~\cite{doyon2017large}).

\subsection{Quantum fluctuations around (non-interacting) Generalized Hydrodynamics}
The goal of the theory of QGHD is to capture long-wavelength quantum fluctuations around the hydrodynamic solution of  the GHD equation (\ref{eq:wigner}). This has been obtained in the free case in~\cite{ruggiero2019conformal}, and later generalized to truly interacting models in~\cite{ruggiero2020quantum}.

\subsubsection{Propagation of quantum fluctuations}

We now briefly recall the main steps for its derivation, restricting to the case of interest, namely the TG limit or, equivalently, non-interacting fermions.
Starting from a zero-entropy hydrodynamic configuration (Eq.~\eqref{eq:wigner_gamma} or, equivalently, Eq.~\eqref{splitFermiseas}), small fluctuations can be expressed as deformations of the contour~\cite{giamarchi2003quantum,cazalilla2004bosonizing}, meaning, locally, of the Fermi points, $p_{a} \to p_{a} + \delta p_{a}$.
As shown in Ref.~\cite{ruggiero2020quantum}, $\{ \delta p_{a} \}$ obey (at first order) the following equation
\begin{equation}\label{eq:soundTG}
	( \partial_t + p_a (x,t) \partial_x ) \delta p_a (x,t) =0,
\end{equation}
describing the propagation of linear sound waves.
%
%
Note that, as an important simplification occurring with respect to the interacting case, here the equations do not couple $p_{a}, p_{b}$ with $a\neq b$. To make contact with Ref.~\cite{ruggiero2020quantum}, this follows from the fact that the \emph{flux jacobian} $A_{a,b}\equiv \partial_a \epsilon /\partial p_b$ with $\epsilon_a = p_a (x, t)^2/2 + V(x,t)$ denoting the semiclassical energy associated to $p_a$, is diagonal in this case (cf. Eq.~(8) there).

One considers the quantized version of such fluctuations, obtained by viewing the small displacement of the contour $\delta p_{a}$ as an operator acting on a Hilbert space, $\delta p_{a} (x,t) \to \delta \hat{p}_{a} (x,t)$, $a=1,\dots, 2Q$. A small displacement of the contour is equivalent to a small excess of particles around point $(x_t,p_t)$ in phase space, so we define the excess density operator as $\delta \hat{\rho}_a (x,t) = \frac{(-1)^a}{2\pi \hbar} \delta \hat{p}_a (x,t)$. Then, using the parameterization (\ref{eq:parameter_s}), it is more convenient to work with the excess density along the contour $\Gamma_t$, defined as
\begin{equation} \label{eq:defrhos}
	\delta \hat{\rho} (s,t) \equiv  \left| \frac{d x_t (s)}{ds} \right|  \delta \hat{\rho}_a (x_t(s) ,t) ,
\end{equation}
for the index $a$ such that $p_t(s) = p_a(x_t(s),t)$. Here the Jacobian $|dx_t(s)/ds|$ is included because we want $\delta \hat{\rho}(s)$ to be a density along the contour; in particular we want
\begin{equation}
	\delta \hat{N} \equiv \int_0^{2\pi} \delta \hat{\rho} (s) ds
\end{equation}
to be the operator that measures the excess of particles in the gas with respect to the average number $N$. Notice that $\delta \hat{N} $ must have integer eigenvalues.

Next, we impose that the operator $\delta \hat{\rho} (s)$ satisfies the $\rm U(1)$ current algebra
\begin{equation} \label{eq:U1}
[\delta \hat{\rho} (s), \delta \hat{\rho} (s')] = \frac{1}{2\pi i} \delta' (s-s')  ,
\end{equation}
so $\delta \hat{\rho}(s)$ is a chiral field that lives along the contour $\Gamma_t$.
{
Note that instead of imposing the commutation relations of total densities and velocities, as in \eqref{eq:landau_canonical}, we use the excess densities at different momenta, which are more natural variables in the QGHD context and lead to the commutation relations of a $\rm U(1)$ algebra.
}
Locally, i.e., in a small cell around a fixed $x$, the above commutation relations are written in terms of several chiral components $\delta \hat{\rho}_a (x)$ (as many as the number of local Fermi points $p_{a}(x)$),
\begin{equation}
[ \delta \hat{\rho}_a (x), \delta \hat{\rho}_b (y)]= \frac{(-1)^a}{ 2\pi i} \delta_{ab} \delta ' (x-y) \ .
\end{equation}
The time-dependent Hamiltonian that generates the dynamics of the quantum fluctuations is~\cite{ruggiero2020quantum}
\begin{equation} \label{Hx}
\hat{H} [{\Gamma_t}] =  \pi \hbar \sum_a \int  dx \, (-1)^a {p_{a}(x,t)} \left( {\delta  \hat{\rho}_a (x)} \right)^2  =  \pi  \hbar \int  ds \, \left( \frac{d x_t(s)}{ds} \right)^{-1} {p_t(s)} \left( \delta \hat{\rho}(s) \right)^2 .
\end{equation}
This Hamiltonian depends on time only through its dependence on the contour $\Gamma_t$, and it is chosen so that it reproduces the evolution equation of sound waves (\ref{eq:soundTG}). Indeed, plugging the Hamiltonian (\ref{Hx}) into the Schr\"odinger equation for $\delta \hat{\rho}_a$, i.e. $\partial_t \delta \hat{\rho}_a(x) = \frac{i}{\hbar} \left[ \hat{H}[\Gamma_t], \delta \hat{\rho}_a(x) \right] $, gives back the wave equation (\ref{eq:soundTG}).
Eq.~\eqref{Hx} is a special case of the QGHD Hamiltonian given in~\cite{ruggiero2020quantum} (for the Tonks-Girardeau gas, the flux Jacobian is $A_{a,b} = \delta_{ab} \, k_{a} $, see the comment above).
For the Tonks-Girardeau gas, time-evolution under the Hamiltonian \eqref{Hx} simply induces a rotation of the field $\delta \hat{\rho}(s)$ together with the contour $\Gamma_t$. Indeed, the Heisenberg evolution equation for $\delta \hat{\rho}(s,t)$ reads (see Eq.~(\ref{eq:defrhos}))
\begin{eqnarray}
\nonumber	\frac{d}{dt} \delta \hat{\rho} (s,t) & = &  \left( \partial_t \left| \frac{d x_t (s)}{ds} \right| \right)  \delta \hat{\rho}_a (x_t(s) ,t)   + \left| \frac{d x_t}{ds} \right|  (\partial_t x_t (s) )  \partial_x \delta \hat{\rho}_a (x_t(s),t) +  \frac{i}{\hbar} [ \hat{H} [\Gamma_t],  \delta \hat{\rho} (s,t) ]  \\ 
\nonumber & = &  {\rm sign}\left( \frac{dx_t}{ ds} \right)  \frac{d p_t(s)}{ds}   \delta \hat{\rho}_a (x_t(s) ,t)  +  \left| \frac{d x_t}{ds} \right| p_t(s) \partial_x \delta \hat{\rho}_a (x_t(s) ,t)   + \frac{i}{\hbar} [ \hat{H} [\Gamma_t],  \delta \hat{\rho} (s,t) ]  \\
\nonumber & = &  {\rm sign}\left( \frac{dx_t}{ ds} \right)  \frac{d p_t(s)}{ds}   \delta \hat{\rho}_a (x_t(s) ,t)  +  {\rm sign} \left( \frac{d x_t}{ds} \right) p_t(s) \partial_s \delta \hat{\rho}_a (x_t(s) ,t)   + \frac{i}{\hbar} [ \hat{H} [\Gamma_t],  \delta \hat{\rho} (s,t) ]  \\
&=&  \frac{\partial}{\partial s} \left(  p_t(s) \left( \frac{dx_t}{ds} \right)^{-1}  \delta \hat{\rho} (s,t) \right) + \frac{i}{\hbar} [ \hat{H} [\Gamma_t], \delta \hat{\rho} (s,t) ]  .
\end{eqnarray}
Note that the above derivation is done assuming to be far from the turning points (this means in particular that the `sign' function is just a constant).
We see that the first term is a convection term, which expresses the fact that the excess density is transported along the contour with a velocity $ p_t(s) \left( \frac{dx_t}{ds} \right)^{-1}$. The second term is fixed by the Hamiltonian (\ref{Hx}); in the general case of interacting theories, it is a non-trivial term, see~\cite{ruggiero2020quantum}. However, for the Tonks-Girardeau gas, it is easy to see from Eq.~(\ref{Hx}) that this term exactly compensates the convection term, so that
\begin{equation} \label{trivial_Eq}
\frac{d}{dt} \delta \hat{\rho} (s,t) =0 \ .
\end{equation}
This is a consequence of $s$ being a co-moving coordinate in the evolution equation~(\ref{eq:xt_pt}), and of the theory being non-interacting (no other terms come from the commutator, in contrast with the interacting case).
Eq.~\eqref{trivial_Eq} means that the evolution of the field is trivial, so that we only need to compute correlations at $t=0$ and the latter are then just ``transported'' in time according to \eqref{eq:evo} [in the general case the same would be ``transported'' according to GHD].

\subsubsection{Quantum fluctuations and correlations at time $t=0$.}

In the previous subsection, we have seen that quantum fluctuations propagate in a very simple way in the Tonks-Girardeau gas: they simply follow the motion of the contour $\Gamma_t$, as expressed by Eq.~(\ref{trivial_Eq}). If we know the correlation functions of the field $\delta \hat{\rho}(s)$ at time zero, then it is trivial to propagate them to later times thanks to Eq.~(\ref{trivial_Eq}).

To find correlation functions at time $t=0$, we observe that the system at $t=0$ must be in the ground state of the Hamiltonian $H[\Gamma_{0}]$. Notice that, so far, we have not made any specific choice of parameterization for the contour, see Eq.~(\ref{eq:parameter_s}). All the equations we wrote so far were valid for an arbitrary parameterization. However, now it is useful to make the following choice, which greatly simplifies all calculations: we choose~\cite{dubail2017conformal,ruggiero2019conformal}
\begin{equation} \label{parameter}
s (x_0)= \pi \frac{\int^{x_0}_{-R/2} dx / p(x,0) }{\int_{-R/2}^{R/2} dx / p(x,0)} \ ,
\end{equation}
where $p (x,t)=  \frac{1}{\pi} \sqrt{2 (\mu - V (x, t))}$ and $[-R/2, R/2]$ is the interval where $V (x, 0)<\mu$. This provides a coordinate $s \in [0, \pi]$ for the upper part of the Fermi contour at $t=0$, which can be continued to $s \in [\pi, 2 \pi]$ to parameterize also the lower part. With this choice, we have
\begin{equation}
	\left( \frac{d x_0(s)}{ds} \right)^{-1} p_0(s) = {\rm constant},
\end{equation}
so that the Hamiltonian at time $t=0$ is simply
\begin{equation}\label{eq:ham_t0}
	H[\Gamma_0] \propto \int_0^{2\pi} (\delta \hat{\rho}(s))^2 ds .
\end{equation}
We recognize the Hamiltonian of a chiral ${\rm U}(1)$ CFT on a circle of circumference $2\pi$. All correlation functions in that theory can be expressed in terms of the ones of the dimensionless bosonic field $\hat{\phi}(s) \in \mathbb{R}/(2\pi \mathbb{Z})$, defined such that
\begin{equation}
	\delta \hat{\rho} (s) = \frac{1}{2\pi} \partial \hat{\phi} (s).
\end{equation}
The operator $\hat{\phi} (s)$ satisfies the $\rm U(1)$ current algebra $[\partial \hat{\phi} (s), \partial \hat{\phi} (s')] = -2\pi i \delta' (s-s') $ as a consequence of (\ref{eq:U1}). The two-point function in the ground state of the quadratic Hamiltonian (\ref{eq:ham_t0}) is
\begin{equation} \label{twopoint}
\langle \hat{\phi} (s) \hat{\phi} (s')  \rangle = - \log \left| 2 \sin \frac{s-s'}{2} \right| ,
\end{equation}
and all other correlation functions of primary operators in the theory can be obtained using Wick's theorem.
We stress that the fact that we have arrived at a CFT is specific to the GHD description of {\it non-interacting} particles~\cite{dubail2017conformal,ruggiero2019conformal}. As an important consequence, the (quadratic) theory of quantum fluctuations around GHD is complemented by conformal symmetry and leads to explicit results for correlation functions. In the interacting case, even though the theory is still quadratic, the Hamiltonian typically involves terms that break conformal invariance~\cite{brun2018inhomogeneous,granet2019inhomogeneous,gluza2021breaking,scopa2020one} (we note that, even in the interacting case, there can exist peculiar situations where conformal invariance survives~\cite{collura2020domain}).
In the general case, the two-point function \eqref{twopoint} needs to be computed numerically, as the solution of a generalized Poisson equation~\cite{brun2018inhomogeneous}.


%
%

\vspace{0.3cm}

Finally, note that one needs to relate observables defined in the initial microscopic model to quantities in the theory.
Specifically, a given observable $O(x, t)$ will take the form an expansion in series~\cite{brun2017one,brun2018inhomogeneous}
\begin{equation} \label{expansion}
\hat{O}(x,t) = \sum_j c_j \hat{\Phi}_j (x,t)
\end{equation}
where, in our case, $\hat\Phi_j (x,t)$ are (conformally normalised) operators of a CFT that we can order according to their scaling dimension, and $c_j$ are non-universal constants, which in inhomogeneous and out-of-equilibrium settings, as the one we consider in this work, will further acquire a space and time dependence, i.e., $c_j \to c_j (x, t)$.
The operators entering the expansion \eqref{expansion} are fixed by transformation under symmetries, dimension requirements, and so on (this was explicitly done, e.g., in~\cite{brun2018inhomogeneous}).

%

%

\begin{figure*}[t]
\centering
\includegraphics[width = 0.93\textwidth]{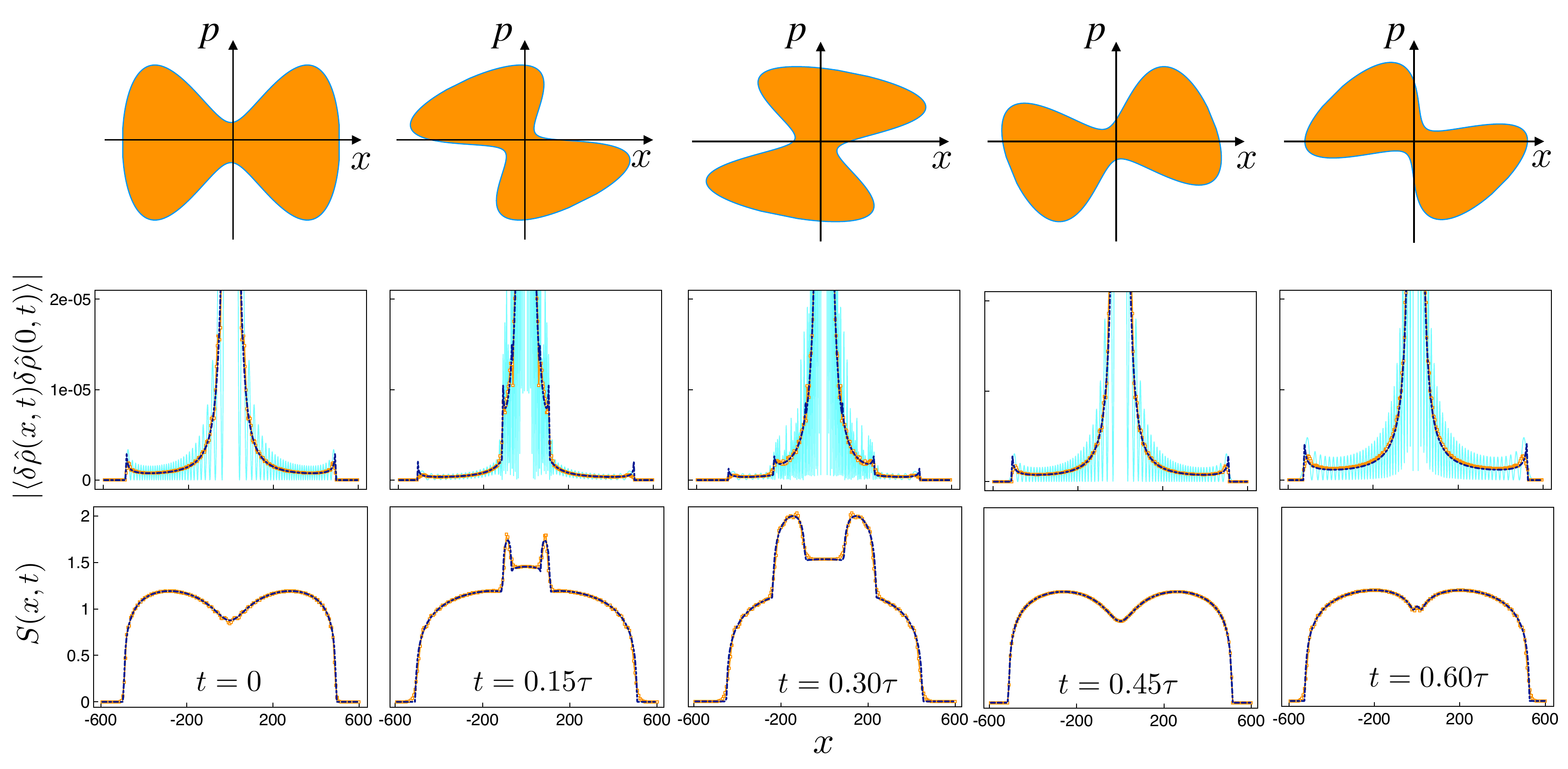}
\caption{Quantum quench from double to single well in the Tonks-Girardeau gas $\bar{g} \rightarrow \infty$. First row: Fermi contour. Second row: Absolute value of the connected density-density correlator. Last row: Entanglement entropy. Each row shows the corresponding quantity as a function of the spatial coordinate $x$, at different times, expressed as a fraction of the period $\tau$ (from $t=0$ in the first column to $t=0.60 \tau$ in the last). 
Orange symbols are the numerical data (obtained from a lattice model in the dilute limit), whereas dashed blue lines are the QGHD predictions. 
The numerics for the density-density correlator shows large oscillations (cyan continuous line), therefore the analytic prediction is compared with its spatial averaging. 
It diverges at coinciding points as $x^{-2}$. 
The parameters are chosen as follows: system size $L=1200$; number of particles $N=66$; pre-quench potential $V(x, t=0)= a_4 x^4 - a_2 x^2$ with $a_4= 6/L^4, a_2= 1/L^2$ and $\mu=0.003$ for the chemical potential; post-quench potential $V(x, t>0) = \omega^2 x^2 /2$ with $\omega=1/L$ (and period $\tau= 2\pi/\omega$).}
\label{movie}
\end{figure*}

\section{Results for the density fluctuations}
\label{sec:results_density}

In this section, we explicitly compute the {leading asymptotic of the} equal-time density-density (connected) correlation function, which {cannot be predicted by GHD alone as  Euler scaling gives simply zero (all correlations within different fluid cells vanish faster than the inverse distance).}

Following the program outlined above, we first express the density operator $\hat{\rho} (x, t)$ in terms of operators in the CFT \eqref{Hx}. Assuming to start with a zero-entropy state (cf.~\eqref{splitFermiseas}), let $x$ and $t$ be such that the corresponding Wigner function $n (x, p, t)$ is associated with $Q$ Fermi seas. 
They can be traced back to their initial positions at $t=0$ via Eq.~\eqref{eq:evo} so that, at initial time, they correspond to the $2Q$ points $\{s_a\}_{a = 1,\cdots, 2Q}$ on the initial contour $\Gamma_0$. 
Then, the claim is that $\hat{\rho} (x, t)$, at leading order, takes the form
\begin{equation}
\hat{\rho} (x,t) = \rho_{\rm GHD} (x, t) + \delta \hat{\rho} (x, t) + \cdots, \qquad 
	\delta \hat{\rho} (x,t) \, = \, \sum_{a=1}^{2Q} \frac{1}{2\pi} \left( \frac{d s_a}{d x} \right) \partial \hat\phi(s_a) 
\end{equation}
where $\rho_{\rm GHD} (x, t) =  \int  \frac{dp}{2\pi \hbar} \, n (x, p , t)$ is the particle density, and the ellipses denote further contributions coming from operators with higher scaling dimension (giving rise to subleading terms in the CFT).
The operator $\delta \hat{\rho}$ measures the fluctuations of the particle density and we are interested in its 
two-point function at time $t$, namely,
\begin{eqnarray} \label{rhorho}
	\left< \delta \hat{\rho} (x,t) \delta \hat{\rho} (x',t) \right>  =   \sum_{a, b=1}^{2Q} \frac{1}{(2\pi)^2} \left( \frac{d s_a}{d x} \right)  \left( \frac{d s_b}{d x'} \right) \langle  \partial \phi(s_a)  \partial \phi(s_b) \rangle \ .
\end{eqnarray}
Each correlator in the sum in the r.h.s. is readily determined: indeed it corresponds to the two point function of the primary field $\partial \hat\phi (s)$ in the ground state of the CFT~\eqref{Hx}, simply given by~\cite{yellowbook}
\begin{equation} \label{dphidphi}
\langle \partial \hat\phi (s_a) \partial \hat\phi (s_b) \rangle=  \frac{1}{\left| 2 \sin \frac{s_a - s_b}{2} \right|^{2 } }.
\end{equation}
%

In the specific protocol depicted in Section~\ref{sec:model}, two regimes are expected: if, at time $t$, there are only two Fermi points at position $x$, then the sums in \eqref{rhorho} stop at $Q=1$; if there are four Fermi points, then, it stops at $Q=2$. With this further prescription, Eq.~\eqref{rhorho} together with Eq.~\eqref{dphidphi} is the final result.

Our prediction is plotted in the second row of Fig.~\ref{movie} and compared with exact numerics, which is achieved by taking advantage of the free fermionic nature of the TG limit (see Appendix~\ref{app:numerics} for details on the implementation) which makes it possible to consider large numbers of particles and long times.
The exact microscopic solution has large Friedel oscillations (cyan curve) -- coming from the presence of the trap, that breaks translational invariance~\cite{cazalilla2002low,das2020friedel,artemenko2004friedel} --, which can be suppressed by spatially averaging over a small window $[x-\Delta x/2,x+\Delta x/2]$. After averaging, the agreement with the prediction of QGHD is remarkable.

One interesting physical feature already noted in Ref.~\cite{ruggiero2020quantum} is the divergence appearing at the points where a change in the number of Fermi points occurs, a genuine predictions of our theory.
While in~\cite{ruggiero2020quantum} such peaks were not visible in the microscopics of the interacting case, 
here (in the TG limit), instead, those are in fact visible also in the averaged microscopic simulations.
Note that, in this case, we are able to get closer to the thermodynamic limit ($N=66$ in the Fig.~\ref{movie} versus $N \le 20$ in Ref.~\cite{ruggiero2020quantum}), and therefore fluid cells are on smaller scale, so that peaks remain as meaningful non-microscopic features, beyond Euler cells.


\section{Result for the entanglement entropy} \label{EE}
\label{sec:results_entanglement}

In this section, we compute the entanglement entropy, as a function of space and time, for the protocol detailed in Section~\ref{sec:model}. While it is known to vanish in GHD \cite{alba2019towards,entGHDfree,entGHDint}, we are going to show that it is instead non-zero in QGHD. 
We also remind that this quantity is particularly challenging to compute directly within the microscopic model and therefore its calculation in this framework manifests the predictive power of our approach. 

The entanglement entropy $S(x, t)$ between the subsystem $A=[-\infty, x]$ and its complement is defined as 
 \begin{equation} \label{entent}
 S(x, t) = -\textrm{tr} \left( \sigma_A \ln \sigma_A \right)
 \end{equation}
 where $\sigma_A$ is the reduced density matrix associated with $A$.
In field theory, $S$ is usually obtained from the corresponding R\'enyi entropies $S_{\alpha} = \frac{1}{1- \alpha} \ln \textrm{tr}\left(\sigma_A^{\alpha}\right)$ (labelled by $\alpha \in \mathbb{R}$) via replica limit ($\alpha \to 1$).
$S_{\alpha}$ can be expressed as the sum of two parts  

%
\begin{equation} \label{renyi_sum}
S_{\alpha} (x, t) = \mathcal{S}_{\alpha} (x, t) - \epsilon (x, t).
\end{equation}
Here $\mathcal{S}_{\alpha} (x, t)$ may be predicted by QGHD complemented with conformal invariance, whereas $\epsilon (x, t)$ is an ultraviolet cutoff. 
In the inhomogeneous situation, such cutoff also acquires a dependence on space and time~\cite{dubail2017conformal}.

The final result, that we are going to derive below, reads
\begin{equation} \label{finalS}
S_{\alpha} (x, t) = -  \frac{1}{12} \left(1 + \frac{1}{\alpha} \right) \left\{ \sum^{2Q}_{a > b} (-1)^{a+b}  \ln \left| \left( p_a (x, t) - p_b (x, t) \right) \sin \frac{s_a - s_b}{2} \right| + \sum_a^{2Q} \ln  \left| \frac{d s_a}{dx} \right|     \right\}  + \Phi_\alpha \ .
\end{equation}
where, depending on the value of $x$ and $t$, $Q=1$ or $2$, and
\begin{equation} \label{Phi_alpha}
\Phi_\alpha = \left( \Upsilon_\alpha + \frac{1}{6} \left( 1+ \frac{1}{\alpha} \right) \log 2 \right) \times \frac{Q}{2}
\end{equation}
is a non-universal constant, with $\Upsilon_1 \simeq 0.4950179 \dots $ for the von Neumann entropy ($\alpha=1$) and the expression for  generic $\alpha$ can be found in~\cite{jk-04}.

The comparison with numerics is shown in the last row of Fig.~\ref{movie}: the agreement is impressive.

\subsection{Universal contribution} \label{sec:uni}

Let us start with the derivation of the QGHD contribution. When $\alpha \in \mathbb{N}$, $\mathcal{S}_{\alpha}$ can be expressed as the correlation function of special fields known as \emph{twist fields} \cite{cc-04,DCC,cc-09}, $\mathcal{T}$ and $\bar{\mathcal{T}}$, local operators lying at the boundary of the subsystem (in our case, the point $x$). Explicitly, we can write 
\begin{equation} \label{Shat}
\mathcal{S}_{\alpha} (x, t) = \frac{1}{1- \alpha} \ln \langle \mathcal{T} (x, t) \rangle.
\end{equation}
In our chiral theory, twist fields are products of chiral excitations. 
We refer to them as \emph{chiral twist fields}  $\mathcal{\tau}$ and $\bar{\mathcal{\tau}}$. 
Crucially, in CFT, they behave as primary fields, with scaling dimension $\Delta= \frac{1}{24} \left( \alpha - \frac{1}{\alpha} \right)$ \cite{cc-04}.
As in the previous example, two regimes are expected: if, at time $t$, there are only two Fermi points at position $x$, then $ \langle \mathcal{T} (x, t) \rangle $ is a two-point function in the chiral CFT that lives along the Fermi contour; if there are four Fermi points $ \langle \mathcal{T} (x, t) \rangle $ is a four-point function.


Let $x$ and $t$ be such that there are two Fermi point. Then, they can be traced back to their initial positions at $t=0$ via Eq.~\eqref{eq:evo}. Denoting the two initial coordinates along the Fermi contour by $s_1, s_2$, Eq.~\eqref{Shat} becomes
\begin{equation} \label{S2a}
\mathcal{S}_{\alpha} = \frac{1}{1-\alpha} \ln \left( \left| \frac{d s_1}{dx} \right|^{\Delta}  \left| \frac{d s_2}{dx} \right|^{\Delta} \langle \tau (s_1) \bar{\tau} (s_2) \rangle \right),
\end{equation}
and
\begin{equation} \label{S2b}
\langle \tau (s_1) \bar{\tau} (s_2) \rangle \simeq \frac{1}{\left| \sin \frac{s_1 - s_2}{2} \right|^{2\Delta} }.
\end{equation}
up to a normalization constant, giving rise to a subleading contribution in the final result (it can actually be computed as well, see Eq.~\eqref{Phi_alpha}).

When at position $x$ and time $t$ we have four Fermi points, they can be traced back to positions $s_1, s_2, s_3, s_4$ along the contour at time $t=0$. 
Then Eq.~\eqref{Shat} becomes
\begin{equation}\label{S4a}
\mathcal{S}_{\alpha} =  \frac{1}{1-\alpha} \ln \left( 
\prod_{a=1}^4  \left| \frac{d s_a}{dx} \right|^{\Delta}
\langle \tau (s_1) \bar{\tau} (s_2)  \tau (s_3) \bar{\tau} (s_4) \rangle
\right),
\end{equation}
where, similarly, 
\begin{equation} \label{S4b}
\langle \tau (s_1) \bar{\tau} (s_2)  \tau (s_3) \bar{\tau} (s_4) \rangle \simeq 
\frac{\left|  \sin \frac{s_1 - s_3}{2} \right|^{2\Delta} \left|  \sin \frac{s_2 - s_4}{2} \right|^{2\Delta}}{\left|  \sin \frac{s_1 - s_2}{2} \right|^{2\Delta} \left|  \sin \frac{s_3 - s_4}{2} \right|^{2\Delta} \left|  \sin \frac{s_1 - s_4}{2} \right|^{2\Delta} \left|  \sin \frac{s_2 - s_3}{2} \right|^{2\Delta}}.
\end{equation}

\subsection{Inhomogeneous cutoff and constant term via a generalized Fisher-Hartwig conjecture}
\label{subsec:fisherhartwig}


The cutoff part $\epsilon (x, t)$ instead is not predicted by conformal invariance and has to be determined by other means, as well as the non-universal constant $\Phi_{\alpha}$ in \eqref{Phi_alpha}.
To do that, we start from a free fermionic model defined on a lattice and consider the continuous limit only at the end.
Taking advantage of the free nature of the problem, we can rely on the Fisher-Hartwig conjecture~\cite{BasTr,Basor1979,jk-04,ce-10}.

Let us start with a uniform system and consider the discrete version of a split Fermi sea, identified by a number of (adimensional) Fermi points $\{ \kappa_a\}$. Those are related to their continuous analog $\{ p_a \}$ by $p_a = \lim_{\delta \to 0} \kappa_a / \delta $, where $\delta$ is the lattice spacing.
The cutoff is a function of them, that we denote as $\epsilon ( \{ \kappa_a \})$.
The expression for two Fermi points is well-known~\cite{jk-04} and can be generalized to a generic number of Fermi seas.
%

The starting point is to write the entanglement entropy as the contour integral
\begin{equation} \label{Sintegral}
S_\alpha = \frac{1}{2 \pi i} \int_{\mathcal{C}} e_\alpha(\lambda) \frac{ d \ln D (\lambda)}{d \lambda},
\end{equation}
where $e_\alpha(x)= \frac1{1-\alpha} \log (x^\alpha+(1-x)^\alpha)$ and  $D(\lambda)= \det ( \lambda -   C_A)$. 
$C_A$ is the correlation matrix of the subsystem $A$ in the split Fermi sea state and is a Toeplitz matrix. The symbol of such matrix in the case of multiple Fermi seas reads
\begin{equation}
g (\kappa) = 
\begin{cases}
1 \qquad \textrm{if} \; \kappa \in \; \textrm{Fermi seas,}\\
- 1 \quad \textrm{if} \; \kappa \not\in \; \textrm{Fermi seas.} 
\end{cases}
\end{equation}
Hence the symbol defining $D(\lambda)$ is $\tilde{g} (\kappa) = \lambda - g (\kappa)$. 
If we consider $Q$ Fermi seas, such function has $2Q$ discontinuities, and can be represented as
\begin{equation}
\tilde{g} (\kappa) = \psi (\kappa) \prod_{a=1}^{2Q} t_{\beta_i (\lambda), \kappa_a} (\kappa) ,
\end{equation}
where the points $\{ \kappa_a \}$ correspond to the location of the discontinuities (i.e., the Fermi points), $\psi (\kappa)$ is the same as for a single Fermi sea~\cite{jk-04} and
for $m\in \mathbb N$
\begin{equation}
\beta_{2m} (\lambda)= - \beta_{2m+1} (\lambda) =
\beta (\lambda)   \equiv \frac{1}{2 \pi i} \ln \left( \frac{\lambda +1}{\lambda-1} \right), \quad t_{\beta_a , \kappa_a} (\kappa)= \exp[-i \beta_i(\lambda) (\pi - \kappa+ \kappa_a)].
\end{equation}
The Fisher-Hartwig conjecture provides the asymptotics of $D(\lambda)$.
We are only interested in the additive constant (with respect to the subsystem size) part of $D(\lambda)$, which provides the cutoff.
In particular, we consider the part of the constant term which depends on $\{ \kappa_a \}$, omitting an overall function of $\lambda$. This is
\begin{equation}
 \prod_{1\leq a \neq b \leq 2Q } \left( 1- e^{i (\kappa_a - \kappa_b) } \right)^{\beta_a \beta_b} .
\end{equation}
The logarithmic derivative in \eqref{Sintegral} then gives
\begin{eqnarray}
\frac{d}{d \lambda} \left[ \sum_{a \neq b} \beta_a(\lambda) \beta_b (\lambda) \ln \left( 1 - e^{i (\kappa_a - \kappa_b)}  \right) \right]
= \left( \frac{d \beta (\lambda)^2}{d \lambda} \right) \sum_{a \neq b} (-1)^{a+b} \ln (1- e^{i (\kappa_a - \kappa_b)})
= 4 \beta ' (\lambda) \beta (\lambda) \epsilon (\{ \kappa_a \}),
\end{eqnarray}
where, in the last equality, we factorised the same expression appearing in the case of a single Fermi sea. Hence we can read off the cutoff as
\begin{equation}
\label{cutoff}
\epsilon (\{ \kappa_a \}) = \frac{1}{12} \left(1 + \frac{1}{\alpha} \right) \left[ \frac{1}{2} \sum_{a \neq b} (-1)^{a+b} \ln \left( 1- e^{i (\kappa_a - \kappa_b)} \right)  \right].
\end{equation}
%
One can verify that Eq.~\eqref{cutoff} reduces to the known expressions in the case of a single Fermi sea \cite{jk-04} and the case of many but symmetric Fermi 
seas \cite{km-05,km-long}.

The continuous limit (in terms of $\{ p_a\}$) is obtained by sending $\delta \to 0$ in \eqref{cutoff}, which gives
\begin{equation}
\label{cutoff2}
\epsilon (\{ p_a \}) \simeq \frac{1}{12} \left(1 + \frac{1}{\alpha} \right) \sum_{a > b} (-1)^{a+b}  \ln \left|  {p_a - p_b} \right|
\end{equation}
where we kept only the leading term, and we disregard a term proportional to $\ln \delta$. The latter, indeed does not enter in the final result \eqref{finalS} as it must be exactly compensated by a similar term in the universal contribution computed in Sec.~\ref{sec:uni}. In fact, it is clear that in Eq.~\eqref{S2a} or \eqref{S4a}, in order to have dimensionless quantities inside the logarithm, the jacobians should be multiplied by a length scale (e.g., reasoning on the lattice, $\delta$).

For inhomogeneous systems, then, we just replace the Fermi points in \eqref{cutoff2} with the spacetime dependent ones $p_a \to p_a (x, t)$.

Finally, also the non-universal constant $\Phi_{\alpha}$ in \eqref{Phi_alpha} is similarly deduced from the result in the homogeneous model, still relying on the Fisher-Hartwig conjecture.

\section{Conclusions}

This work is a follow-up of Ref.~\cite{ruggiero2020quantum}.
We consider a 1d Bose gas undergoing the same quench dynamics as in \cite{ruggiero2020quantum}, but focusing on the Tonks-Girardeau (infinite repulsion) limit. In this case, on the analytical side, more explicit results can be obtained by taking advantage of the restored conformal invariance. On the numerical side more stringent tests can be performed.
In particular, analytical results are provided for the density fluctuations and the entanglement entropy after the quench, which are exact at the Euler scale, and systematically checked against numerics. 
For the entanglement entropy a non-universal space and time dependent contribution appears, which we are able to access by a generalization of the Fisher-Hartwig conjecture.

The theory introduced in Ref.~\cite{ruggiero2020quantum} and further analysed in the present work has the potential to be applied to many out-of-equilibrium situation, as the one considered in this paper, and opens several further directions, also in connections with experiments.

A quantity of clear experimental interest is the one-particle density matrix, namely two-point correlation of the boson $\Psi$ entering the Lieb-Liniger Hamiltonian \eqref{model}. This was obtained in \cite{ruggiero2019conformal} for a quench in a single-well potential, after a sudden change of the frequency. The extension of that analysis to the protocol studied here or, more generally speaking, in presence of multiple Fermi points, requires to work out the expression of $\Psi$ in terms of the CFT operators (cf. Eq.~\eqref{expansion}), something that {we plan to address in the near future. In the TG limit, this is something that would be experimentally accessible with modern setups such as the ones in Refs.~\cite{wilson2020observation,malvania2020generalized}.} The case of a finite interaction would be also interesting to investigate.

Moreover, we stress that the quantum GHD theory is defined when starting from zero-entropy states (cf. Eq.~\eqref{eq:wigner_gamma}). While they naturally arise in the thermodynamic limit of ground states of trapped bosons, this is not the case for thermal states, and how to extend the theory in this case is not straightforward. Indeed, by introducing a finite temperature, the Fermi contour entering the definition of zero-entropy Wigner functions gets smoothen, which in turn makes it difficult to clearly define chiral excitation around Fermi points. Still, this should be doable, at least at low temperature (similarly to what happens for standard Luttinger Liquid theory), and it would be interesting to study the interplay between thermal and quantum fluctuations.

A final point, mentioned in Section~\ref{sec:QGHD_TG}, is that Eq.~\eqref{eq:wigner} for the evolution of the Wigner function is only valid at leading order when the potential $V(x, t)$ is not quadratic. 
Otherwise, corrections to that come as a series expansion in $1/\hbar$, also known as Moyal expansion~\cite{moyal1949stochastic}. Such corrections have been recently analyzed in great details~\cite{fagotti2017higher,fagotti2020locally} for free models, while for now it is not understood how to include them in truly interacting systems.
We stress that they are different from the ones we are interested in, and in our case they are actually not there (the post-quench potential is quadratic). In general, however, the relation between the two needs to be clarified: eventually, taking into account the whole Moyal series should correspond to solving exactly the original microscopic model, thus including the `quantum' effects we can describe with the quantum GHD approach.


\acknowledgements

We would like to thank Stefano Scopa for useful discussions and collaboration on closely related topics.
This work has been supported by  ANR through Project ANR QUADY (ANR-20-CE30-0017-02) and CNRS through the Emerging International Actions under the grant QuDOD (JD), by the Swiss National Science Foundation under Division II (PR), by the ERC under Consolidator grant  number 771536 NEMO (PC).

\appendix

\section{Details of the numerics}
\label{app:numerics}

In order to simulate the Tonks-Girardeau limit of the Lieb-Liniger model~\eqref{model}, we consider its fermionic analog~\eqref{Hf} and discretize it on a lattice.
As well known, the procedure to go from a continuous to a discrete model is not unique. 
In our case, we focus on the following discretized quadratic Hamiltonian
\begin{equation} \label{Hdiscrete}
H_d (t) =  \sum_{i=1}^{L-1} J^{(1)}_i (c_{i}^{\dag} c_{i+1} + c_{i+1}^{\dag} c_i) + \sum_{i=1}^{L-2}  J^{(2)}_i (c_{i}^{\dag} c_{i+2} + c_{i+2}^{\dag} c_i) + \sum_{i=1}^L V_i (t) c_i^{\dag} c_i
\end{equation}
which includes next-nearest neighbour hopping. Above, $c_i$ are spinless fermionic operators satisfying the canonical anticommutation relations $\{c_i, c_j^{\dagger} \} = \delta_{ij}$.
In \eqref{Hdiscrete} $L$ is the system size (that, due to the inhomogeneity, we need to keep finite to diagonalize the problem numerically), $J_i^{(1)}$ and $J_i^{(2)}$ are (inhomogeneous) hopping amplitudes, and $V_i (t)$ is the trapping potential.
For our quench protocol, the latter is given by 
\begin{equation}
V_i (t) = 
\begin{cases}
a_4 \left( i - L/2 \right)^2 - a_2 \left( i -L/2 \right)^2 & t=0 \\
\omega \left( i -L/2 \right)^2  &t>0
\end{cases} \ .
\end{equation}
For the simulations in Fig.~\ref{movie} the parameters are fixed as follows: $J^{(1)} = -2/3, \, J^{(2)} = 1/24, \, a_4 = 6/L^4, \, a_2 = 1/L^2, \omega = 1/L $ and the chemical potential is fixed to $\mu = 0.003$ (corresponding to $N=66$ particles in the ground state of the double well potential).
We checked that, with this choice, the curves for the density profile after the quench looks perfectly periodic (namely, they are indistinguishable at $t=0$ and $t=\pi/\omega$).

To access the quantities of interest, we only need to diagonalize the two-point correlation function
\begin{equation}
C_{ij} (t) = \langle c_i^{\dag} (t) c_j (t) \rangle  \ ,
\end{equation}
where $c_{i}^{(\dag)} (t) = U^{\dag} (t) c_i U(t) $ are the time evolved creation/annihilation operators (with $U(t) =e^{i H_d (t>0) t}$ the time-evolution operator).

Indeed, the state of the system is gaussian at all times, so that one can rely on Wick theorem~\cite{wich1950evaluation} to compute the (connected) density-density correlations as
\begin{equation}
\langle \delta \rho_i (t) \delta \rho_j (t) \rangle 
=  C_{ij} (t) ( \delta_{ij} - C_{ji} (t) )\ ,
\end{equation}
where $\delta \rho_i (t)= \rho_i (t) - \langle \rho_i(t) \rangle $, and $\rho_i (t) = c_i^{\dag} (t) c_i (t)$ is the local discrete density.

Moreover, for gaussian states, standard free fermions techniques can be used to compute the entanglement entropy~\cite{peschel1999densitymatrix,peschel1999density,chung2001density,peschel2003calculation,peschel2004reduced,peschel2009reduced}. Specifically, if for the subsystem $A= [- L , i]$ of the whole system, we define the two-point function restricted to $A$ with matrix elements $C^A_{ij} (t) = C_{ij} (t)$ for $i,j \in A$, and $\{ \nu_j \}$ is its spectrum, then the entanglement entropy between $A$ and the rest (cf. Eq.~\eqref{entent}), is given by
%
\begin{equation}
S (i, t) =  - \sum_j \left( \nu_j \log \nu_j + (1- \nu_j) \log (1-\nu_j) \right) \ .
\end{equation}

The two-point function, in turn, can be computed via the following standard procedure. 
First, one needs to obtain the single-particle eigenstate amplitude of the pre-quench and post-quench Hamiltonian.
By rewriting \eqref{Hdiscrete} as
\begin{equation}
H_d (t) = \sum_{ij} c_i^{\dag} \,  \mathbb{H}_{ij} (t) \, c_j \,
\end{equation}
this is obtained by diagonalizing the $L \times L$ matrices $\mathbb{H} (t=0)$ and $\mathbb{H} (t>0)$, respectively. 
If we denote by $|\eta_k^0 \rangle$ and $|\eta_q \rangle $, respectively, the eigenstates of such matrices, the two sets of corresponding eigenstates amplitudes are given by $ \eta^0 (k, i) \equiv \langle c_i | \eta^0_k \rangle  $ and $\eta (q, i) \equiv \langle c_i | \eta_q \rangle $, with eigenvalues $\{ \epsilon^0_k \}$ and $\{ \epsilon_q \}$, for pre- and post-quench.
$C_{ij} (t)$ is then expressed in terms of such quantities only. 
In fact, using the explicit form of the ground state of $\mathbb{H} (t=0)$
\begin{equation}
|\psi_0 \rangle = \eta_{k_N}^{0 \, \dag} \eta_{k_{N-1}}^{0\, \dag} \cdots \eta_{k_1}^{0 \, \dag} |0 \rangle
\end{equation}
with $|0 \rangle$ the state annihilated by fermionic operators $\eta^{0}_k \, ( \forall k)$ -- those bringing $\mathbb{H} (t=0)$ in diagonal form --, and the decomposition
\begin{equation}
c_i (t) = \sum_q \eta (q, i) \, \eta_q (t), \quad \eta_q (t) = \eta_q \, e^{-i \epsilon_q t}
\end{equation}
it can written as
%
%
%
%
\begin{equation}
C_{ij} (t) = \sum_{q, p =1}^L \sum_{l=1}^N e^{i t (\epsilon_q - \epsilon_p)}
\,  \eta^* (q,i) \,  \eta (p,j) \, 
\alpha_{k_l}^q
\alpha_{k_l}^{p \, *}, \qquad \alpha_{k}^q   \equiv \langle \eta^0_k | \eta_q \rangle
= \sum_{n=1}^L  \eta^{0 \, *} (k,n) \,  \eta (q, n) 
\end{equation}
which is straightforwardly implemented.

%
\bibliography{qGHD}

\end{document}